\documentstyle[11pt,newpasp,twoside,epsf]{article}
\def\deg{$^{\rm o}$}

\def\kms{\ifmmode {\rm km\ s}^{-1} \else km s$^{-1}$\fi}
\def\Msun{\ifmmode M_{\odot} \else $M_{\odot}$\fi}
\def\Lsun{\ifmmode L_{\odot} \else $L_{\odot}$\fi}
\def\Msigma{\ifmmode M\mbox{--}\sigma \else $M$--$\sigma$\fi}

\def\vFWHM{\ifmmode v_{\mbox{\tiny FWHM}} \else
            $v_{\mbox{\tiny FWHM}}$\fi}
\def\Halpha{\ifmmode {\rm H}\alpha \else H$\alpha$\fi}
\def\Hbeta{\ifmmode {\rm H}\beta \else H$\beta$\fi}
\def\Hgamma{\ifmmode {\rm H}\gamma \else H$\gamma$\fi}
\def\Hdelta{\ifmmode {\rm H}\delta \else H$\delta$\fi}
\def\Lya{\ifmmode {\rm Ly}\alpha \else Ly$\alpha$\fi}
\def\Lyb{\ifmmode {\rm Ly}\beta \else Ly$\beta$\fi}

\def\ciii{\ifmmode {\rm C}\,{\sc iii} \else C\,{\sc iii}\fi}
\def\civ{\ifmmode {\rm C}\,{\sc iv} \else C\,{\sc iv}\fi}

\def\o5007{[O\,{\sc iii}]\,$\lambda5007$}

\begin{document}

\title{Masses of Black Holes in Active Galactic Nuclei} 
\author{Bradley M. Peterson}
\affil{Department of Astronomy, The Ohio State University,
140 West 18th Avenue, Columbus, OH 43210, USA}
\begin{abstract}
We present a progress report on a project whose goal
is to improve both the precision and accuracy of
reverberation-based black-hole masses. Reverberation
masses appear to be accurate to a factor of about three,
and the black-hole mass/bulge velocity dispersion
($\Msigma$) relationship appears to be the same in
active and quiescent galaxies.
\end{abstract}

\section{The Importance of Reverberation Masses}
A detailed knowledge of the masses of supermassive black holes (SBHs) 
in AGNs is an important element of our understanding of how
the accretion process works in nature. Highly accurate (to 10\% or so)
masses are needed to understand the energy budgets in AGNs: how much
of the energy is converted into radiation, how much appears in the
form of kinetic energy in jets and winds, and how much (if any) is
advected in the black hole itself? Less accurate mass determinations
(to within a factor of a few or so) are nevertheless very useful
for current-epoch black-hole demographics that allow
us to approach the same questions about accretion physics from a
statistical point of view. Current estimates of black-hole masses
are anchored by nearly 40 AGNs (Wandel, Peterson, \& Malkan 1999; Kaspi 
et al.\ 2000; Onken et al.\ 2002) whose SBH masses have been
estimated by emission-line reverberation methods (Blandford
\& McKee 1982; Peterson 1993, 2001), in which the response time of the
broad emission lines to continuum variations is taken to be the 
light-travel time across the line-emitting region. By combining this length
scale with a measurement of the line width, a virial mass estimate is
obtained through
\begin{equation}
M = \frac{k c \tau V^2}{G},
\end{equation}
where $c$ and $G$ are the speed of light and gravitational constant,
respectively, $\tau$ is the response time scale for a broad emission
line, $V$ is a suitable measure of the emission-line width, and $k$ is
a numerical factor that depends on the currently unknown structure and
kinematics of the line-emitting region. 
There are two strong pieces of evidence that
suggest that reverberation masses are reliable to within a factor of a
few:
\begin{enumerate}
\item In four AGNs for which multiple emission-line lags and line 
widths have been measured, a virial relationship between
lag and line width (i.e., $\tau \propto V^{-2}$) is seen (Peterson \&
Wandel 1999, 2000; Onken \& Peterson 2002).
\item AGN SBH masses seem to follow the same relationship between
black-hole mass and host galaxy bulge velocity dispersion,
the $\Msigma$ relationship (Ferrarese \& Merritt 2000;
Gebhardt et al.\ 2000), that is seen
in quiescent galaxies (Ferrarase et al.\ 2001, and references therein).
\end{enumerate}
Masses obtained by reverberation methods are becoming increasingly
important as they are used to anchor scaling methods that allow
estimation of the black-hole masses in higher-redshift,
higher-luminosity AGNs (e.g., Wandel, Peterson, \& Malkan 1999;
Vestergaard 2002; McLure \& Jarvis 2002) which thus allows us in
principle to probe the growth of quasar masses over most of the
history of the Universe. Furthermore, unlike stellar-dynamical and
gas-dynamical methods, the reverberation mapping technique is directly
applicable to higher-redshift sources because it is independent of
angular resolution.

Given the potential importance
of the reverberation masses for these sources and the simple fact that
a similar or superior sample is not likely 
to be obtained in the immediate future, we
decided to undertake a detailed reanalysis of the existing
reverberation data to improve on the homogeneity and accuracy
of the database.
In this contribution, we describe some preliminary results.

\section{Refining Reverberation Masses}
Our intent is to improve both the accuracy and precision
to which reverberation masses are known.
The specific goals of this investigation are:
\begin{enumerate}
\item {\em To minimize the formal errors in BLR radius and line width
determinations.} The goal is to increase the measured {\em
precision} of the virial product $rV^2$.  We assume that the virial
relationship holds, and test various characterizations of the
emission-line time lag and velocity width to minimize the scatter in
the virial product for AGNs for which multiple measurements are
available.
\item {\em To determine the best characterization of the
black-hole mass.} The goal is to improve the {\em
accuracy} of reverberation masses by determining the constant $k$ in
eq.\ (1). Here we assume that AGNs and quiescent galaxies follow the
same $\Msigma$ relationship and that $k$ can be determined by scaling
the AGNs relative to quiescent black holes. In sacrificing the chance
of finding systematic differences between active and quiescent
galaxies, we are able to constrain the BLR geometry and kinematics in
a statistical sense by determining both $k$ and its corresponding
uncertainty, i.e., the scatter in the AGN \Msigma\ relationship.
\end{enumerate}

\section{Some Preliminary Results}
As we will describe in detail elsewhere, we conducted tests based on 
both numerical simulations and different treatments of actual
monitoring data. The large database on NGC 5548 (see Peterson et al.\ 2002 
and references therein), for which there are 14 years of 
optical monitoring data plus two years of UV monitoring,
plays a particularly important role in this analysis.
We examined the data carefully to determine
which measures of lag and line width give the most consistent virial
product $rV^2$. We reach the following conclusions:
\begin{enumerate}
\item In general, the centroid of the cross-correlation function is a 
more robust estimate of the lag than the peak value of 
the cross-correlation. 
\item The width of an emission-line in the rms spectrum\footnote{The 
multiple spectra obtained in a reverberation-mapping experiment can be 
used to compute both a mean and root-mean-square spectrum. The latter 
isolates the variable part of the spectrum.} better characterizes the 
line-of-sight velocity dispersion than does the line width in the mean 
spectrum.
\item The line dispersion (i.e., the second moment of the emission-line 
profile) is superior to the full-width at half maximum (FWHM) as a 
measure of the line-of-sight velocity width. However, both 
measures give a slope consistent with $\tau \propto V^{-2}$. 
\end{enumerate}

\section{The {\boldmath $\Msigma$} Relationship for AGNs}
\begin{figure}
\plotfiddle{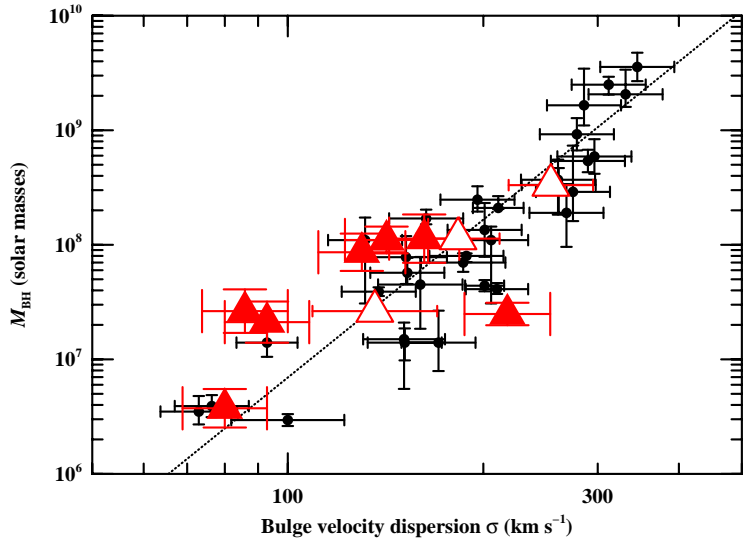}{180pt}{0}{40}{40}{-160}{-27}
\caption{The solid line is the $\Msigma$ relationship
for quiescent galaxies, which are shown as small
filled circles. The large triangles are AGNs from this
work, with $k=9.57$ (see eq.\ 1). The open triangles
are NGC 7469, NGC  5548, and Akn 120 (left to right),
which have the highest precision virial products and
dominate the determination of $k$.}
\end{figure}

If we make the reasonably well-justified assumption that
AGNs follow the same  \Msigma\ relationship as quiescent
galaxies, then we can compute the scale factor $k$ in eq.\ (1)
by scaling the raw virial products $rV^2$ to the 
nominal quiescent-galaxy \Msigma\ relationship. In a statistical sense, 
this can provide us with an admittedly quite weak 
statistical constraint on the structure of the broad-line region. 
The result of such a scaling
exercise is shown in Fig.\ 1, for which we find $k = 9.57 \pm
0.15$. This weighted fit is dominated by Akn 120, NGC 5548, and NGC
7469, the three AGNs with the highest-precision virial products. The scatter of
the AGNs around the $\Msigma$ relationship is about 0.5 dex; i.e.,
reverberation masses appear to be accurate to a factor of about three.

For comparison, we note that the values of $k$ for various
trivial BLR models are $k = 3$ for an isothermal sphere or
for a thin spherical shell of randomly inclined Keplerian orbits, 
and $k = 2/\sin^2 i$ for a ring at inclination $i$. In the
latter case, a typical inclination of $i\approx 27$\deg\
would yield our present estimate of $k$. The fact that
our inferred value of $k$ is in the expected range for
simple models reassures us that our assumption that the
$\Msigma$ relationships for active and quiescent galaxies
are identical is probably not grossly in error.

\acknowledgments
The work described here is being undertaken with my
colleagues 
K.M.\ Gilbert, C.A.\ Onken, R.W.\ Pogge, and M.\ Vestergaard (Ohio State),
L.\ Ferrarese and D.\ Merritt (Rutgers),
S.\ Kaspi, D.\ Maoz, and H.\ Netzer (Tel-Aviv Univ.),
and A.\ Wandel (Hebrew Univ.). We are grateful for 
funding of this work through NASA grant NAG5-8397,
National Science Foundation grant AST-0205964,
and US-Israel Binational Science Foundation grant 1999336.

\end{document}